\documentclass[12pt]{article}

\usepackage{amsfonts}

\begin{document}

\title{How to explain grokking}

\author{S.V. Kozyrev\footnote{Steklov Mathematical Institute of Russian Academy of Sciences, Moscow, Russia, 8 Gubkina St., 119333, {\tt kozyrev@mi-ras.ru}}}

\maketitle

\begin{abstract}
Explanation of grokking (delayed generalization) in learning is given by modeling grokking by the stochastic gradient Langevin dynamics (Brownian motion) and applying the ideas of thermodynamics.
\end{abstract}


\section{Introduction}

Stochastic gradient descent, which is a basic model in machine learning, when taken in the form of the stochastic gradient Langevin dynamics, is a direct analogue of a random walk along the reaction coordinates, which is a principal model in chemical kinetics. It is natural to use in learning the ideas of kinetics and thermodynamics, in particular, if the gradient descent is an analogue of the energy driven chemical reaction, then the stochastic gradient descent should include also the discussion of entropy, i.e. of the free energy minimization. We show that the transition to generalization in grokking is an analogue of the entropy driven chemical reaction and is related to the second law of thermodynamics (see the discussion below).

\section{Learning and thermodynamics}

\subsection{SGLD, free energy and Eyring's formula}

Gradient descent is described by the differential equation
$$
\frac{d x}{dt}=-f'(x).
$$

For the numerical iteration of the descent, the vector $x$ will change as
$$
x_{k+1}=x_{k} - \alpha_k f'(x_{k}).
$$

The stochastic gradient Langevin dynamics (or SGLD) is given by iteration of
$$
x_{k+1}=x_{k} + w_{k} - \alpha_k f'(x_{k}),
$$
where $w_{k}$ are independent Gaussian random vectors.

Continuous limit of SGLD is described by the stochastic differential equation (SDE) for Brownian motion in a potential
\begin{equation}\label{stochastic_equation}
d\xi^i(t)=\sqrt{2\theta}dw^i(t)-{\partial f(\xi(t))\over\partial x^i} dt,
\end{equation}
where $dw^i(t)$ is the stochastic differential of Wiener process which satisfies
$$
dw^i(t)dw^j(t)=\delta_{ij}dt.
$$

This equation was discussed in relation to learning in \cite{Parisi1}, \cite{Parisi2}.

The Fokker--Planck equation for SGLD (\ref{stochastic_equation}) reduces to the diffusion equation in the potential
\begin{equation}\label{diff}
{\partial u\over \partial t}=\theta\Delta u+ \nabla u \cdot \nabla f+ u \Delta f,
\end{equation}
where $x\in\mathbb{R}^d$, $u=u(x,t)$ is the distribution function, $f=f(x)$ is the potential, $f\in C^2(\mathbb{R}^d)$,
$\theta>0$ is the temperature.
Equivalently,
$$
{\partial u\over \partial t}=\theta\,{\rm\bf div}\,\left[e^{-\beta f}\,{\rm\bf grad}\,
\left[u e^{\beta f}\right]\right],
$$
therefore the Gibbs distribution $e^{-\beta f}$, $\beta=1/\theta$ is a stationary solution of the equation, and the solution converges to the Gibbs distribution (under some conditions on $f$).

\medskip

\noindent{\bf Remark}. Let us discuss a short introduction to statistical mechanics for mathematicians, based on arguments from \cite{Cycon}.
Let us put (\ref{diff}) as
$$
{\partial u\over \partial t}=-\theta e^{-\beta f/2} A^{*}A e^{\beta f/2}u,\quad A= e^{-\beta f/2}\nabla e^{\beta f/2}.
$$

The operator $A^{*}A$ is positive, moreover
$$
e^{-t\theta e^{-\beta f/2}A^{*}Ae^{\beta f/2}}=e^{-\beta f/2}e^{-t\theta A^{*}A}e^{\beta f/2},
$$
$e^{-t\theta A^{*}A}$ is a contraction semigroup in $L_2$. The condition $e^{\beta f/2}u\in L_2$ is
\begin{equation}\label{int1}
\int e^{\beta f(x)}u^2(x)dx<\infty.
\end{equation}

For the stationary solution $e^{-\beta f}$ eq. (\ref{int1}) is the integrability of the Gibbs distribution
\begin{equation}\label{int2}
\int e^{-\beta f(x)}dx<\infty,
\end{equation}
zero is the eigenvalue of $A^{*}A$
$$
\left[e^{-\beta f/2}A^{*}Ae^{\beta f/2}\right]e^{-\beta f}=0.
$$

Then if the initial condition $u_0$ satisfies (\ref{int1}) and Gibbs distribution is integrable (\ref{int2}), the dynamics $e^{-t\theta A^{*}A}e^{\beta f/2}u_0$ belongs to $L_2$ and converges to $e^{-\beta f/2}$, and the solution $e^{-\beta f/2}e^{-t\theta A^{*}A}e^{\beta f/2}u_0$ of (\ref{diff}) converges to Gibbs distribution $e^{-\beta f}$, i.e. we get thermalization.

\medskip

\noindent{\bf The Eyring formula of kinetic theory}  describes the reaction rate (for the transition between two potential wells due to diffusion of the type (\ref{diff})): the reaction rate is proportional to
\begin{equation}\label{Eyring}
e^{-\beta(F_1-F_0)},
\end{equation}
where $F_1$ is the free energy of the transition state (the saddle between two potential wells) and $F_0$ is the free energy of the initial state of the reaction (the potential well from which the transition occurs).

The free energy of a domain $U$ (for Gibbs distribution) is
$$
e^{-\beta F(U)}=\int_{U}e^{-\beta E(x)}dx,
$$
if we will take constant energy $E$, we get $F=E-\theta S$, where $E$ is the energy and $S$ is the entropy of the state $U$.

\medskip

Gradient descent in the potential $f$ is the minimization of energy, and stochastic gradient descent is the minimization of free energy.

\subsection{SGLD and learning}

Let a training sample $\{z_l\}$, $l=1,\dots,L$ and a loss function $\mathcal{L}(z,x)\ge 0$ for test $z$ and hypothesis $x$
(let the hypothesis space be $\mathbb{R}^d$) be given.

Minimizing the empirical risk (learning) is the problem
\begin{equation}\label{min}
f(\{z\},x) = \frac{1}{L}\sum\limits_{l=1}^L \mathcal{L}(z_l,x) \,\to\, \min\limits_{x}.
\end{equation}

We will assume $\mathcal{L}(z,x)\ge 0$, $\forall z$, hence $f(\{z\},x)\ge 0$.

Overfitting is the failure to generalize to a learning problem (\ref{min}) when the sample $\{z\}$ is replaced (in particular, low risk for training sample, high risk for validation sample).

We will use the known approach of flat minima:
narrow (sharp) minima of empirical risk (in the hypothesis space) are associated with overfitting, and wide (flat) minima correspond to solutions of the learning problem with generalization \cite{Stab4}.
This known observation is related to algorithmic stability, i.e. stability of the solution of a learning problem to perturbations of the training sample \cite{Stab1}, \cite{Stab2}, \cite{Stab3}.

Learning by stochastic gradient descent for the learning problem (\ref{min}) we will describe by the SGLD (\ref{stochastic_equation}) in the hypothesis space, i.e. for distribution we get the diffusion equation in the potential (\ref{diff}). Therefore, the distribution function $u(x)$ converges to the Gibbs distribution concentrated in potential wells with low free energy.

This predicts the capture of the SGLD learning result by wide potential wells (due to the entropic part of free energy), i.e., the reduction of the overfitting effect in the flat minima approach.
We believe the same effect will work for other forms of stochastic gradient descent (such as mini-batch procedure).
This (and other) approaches to control overfitting are discussed in \cite{Eyring}.

Therefore, we consider the stochastic gradient optimization as a minimization of free energy in statistical mechanics, or an analogue of chemical reaction.
Since entropic contribution to the free energy is important, effects of entropy should be discussed in learning theory.

\section{Grokking and thermodynamics}

Grokking (delayed generalization) phenomenon in learning was discovered in \cite{Grokking1}, see also \cite{Grokking2}, \cite{Omnigrok}, \cite{Grokking3}. Overparameterized network was trained to perform modular arithmetics (in particular, addition, multiplication and other operations modulo 97). The phenomenon of delayed generalization was found, and the following observations were made:

1) The model was trained in $10^3$ steps (by the stochastic gradient descent), memorization of training sample was achieved with almost $100\%$ error on the validation set (i.e. total overfitting). If the training  was continued, then in about $10^6$ steps total generalization was obtained (no overfitting).

2) Threshold behavior was found: grokking works for sufficiently large learning samples.

3) Exponential growth of grokking time with decreasing of training sample was observed.

4) Some kind of principal component analysis for embeddings of residues in the obtained model of modular arithmetics was performed.
It was observed that embeddings of residues lie approximately on a circle, adding a residue is a shift along such a circle. Therefore, modular addition, performed by the neural network, looks like ''structure'' (an algorithm, given by trigonometric polynomials \cite{Grokking3}).

We propose the following explanation of grokking.

For overparameterized models local minima merge into a manifold of zero empirical risk in the hypothesis space \cite{Belkin}, given by the condition (since the loss function is non-negative)
$$
f(\{z\},x)=0,\quad {\rm i.e.}\quad
\mathcal{L}(z,x)= 0,\, \forall z\in \{z\}.
$$

As the training sample $\{z\}$ grows, the zero-risk manifold $f(\{z\},x)=0$ shrinks since additional conditions $\mathcal{L}(z,x)= 0$ for $z\in\{z\}$ are imposed.

The zero-risk manifold contains narrows, or ravines (regions with lower entropy) and wide valleys (with higher entropy).
Ravine landscapes were discussed by Gelfand and Tsetlin, and the ravine method \cite{Gelfand} was developed to improve optimization for such a landscape.

For ravine, or river--valleys landscape, learning works in two steps: the system falls in a ravine, and then performs motion along the ravine.
The generalizing solution for learning problem (''structure'', for example algorithm of modular addition) lies in the high-entropy region of the zero-risk manifold. By the second law of thermodynamics the system performing Brownian motion will travel to regions of high entropy at the zero risk manifold (which is the grokking transition).

Memorization of the training sample is achieving the zero risk manifold $f(\{z\},x)=0$ by the stochastic gradient descent with non-zero gradient, in this case the path traveled in the hypothesis space is proportional to the number of descent steps $\sim t$.

Grokking is a random walk in the zero risk manifold $f(\{z\},x)=0$ (or Brownian motion). For Brownian motion the path traveled is proportional to the root of the number of steps $\sim \sqrt{t}$.

If we assume that the memorization and grokking result in similar path lengths, then the duration of grokking (in the SGD steps) will be the square of the duration of memorization, which is observed in simulation ($10^3$ and $10^6$).

Other properties of grokking can be discussed as follows. The zero-risk manifold contains regions without generalization (i.e. with overfitting) and some region with generalization. Grokking is the transition from the overfitting region to the generalizing (grokking) region of the zero-risk manifold.  Entropy $S_1$ of the region with generalization does not depend on the training sample (by the generalization property), and entropy $S_0$ of overfitting region reduces with the increase of the training sample. This explains the threshold behavior of grokking: for sufficiently large samples the entropy $S_0$ of the overfitting region becomes less than the entropy $S_1$ of the generalizing region, and second law of thermodynamics predicts transition to generalization by increasing of entropy.

Let us suppose that the imposition of each additional condition $\mathcal{L}(z,x)= 0$, $z\in\{z\}$ removes an equal percentage of the volume of the overfitting part of the zero-risk manifold. Then the entropy $S_0$ of the overfitting region decreases linearly as the sample size increases.

The Eyring's formula $e^{-\beta(F_1-F_0)}$ gives an approximation of the dependence of the reciprocal of the grokking time on the training sample size.
Here $F_0$ and $F_1$ are free energies of the overfitting region and of the generalizing region.
The entropy of the overfitting region decreases linearly with the training sample size (hence the free energy $F_0=E_0-\theta S_0$ increases linearly). The free energy $F_1$ of the generalizing region is constant. Consequently, the grokking time decreases exponentially when the training sample increases (this agrees with observations in \cite{Grokking1}).

\medskip

\noindent{\bf Remark}. In \cite{Grokking1}, the AdamW accelerated gradient descent algorithm was used for optimization. At the same time, in a zero-risk manifold, the gradient vanishes, and AdamW turns into a random walk without acceleration. When hitting the walls of ravines, the acceleration of the gradient can affect the dynamics, but this does not affect the progress along the ravine, which is well described by Brownian motion, as can be seen from the comparison of the properties of grokking and Brownian motion described above.

\section{Summary}

Simple ideas of thermodynamics and kinetic theory allow us to explain the grokking (delayed generalization) phenomenon and some properties of grokking observed in \cite{Grokking1} as a manifestation of thermodynamic and kinetic phenomena --- transition to generalization in grokking proceeds by the second law of thermodynamics, of by entropic mechanism of chemical reaction, properties of grokking are explained by Brownian motion and Eyring's formula. Delay of generalization in grokking is a result of transition through the barrier of entropy in Eyring's formula.

\bigskip

\noindent{\bf Acknowledgments}

This work was supported by the Russian Science Foundation under grant no.24-11-00039, https://rscf.ru/en/project/24-11-00039/ .

\end{document}